%% file: better_personalized_medicine.tex
\documentclass[12pt]{article}

\include{preamble}
\title{Evaluating the Effectiveness of Personalized Medicine with Software}

\author[1]{Adam Kapelner\thanks{Electronic address: \texttt{kapelner@qc.cuny.edu}; Principal Corresponding author}}
\author[2]{Justin Bleich}
\author[1]{Alina Levine}
\author[3]{Zachary D. Cohen}
\author[3]{Robert J. DeRubeis}
\author[2]{Richard Berk}

\affil[1]{\small Department of Mathematics, Queens College, CUNY}
\affil[2]{Department of Statistics, The Wharton School of the University of Pennsylvania}
\affil[3]{Department of Psychology, University of Pennsylvania}

\begin{document}
\maketitle

\begin{abstract}
We present methodological advances in understanding the effectiveness of personalized medicine models and supply easy-to-use open-source software. Personalized medicine involves the systematic use of individual patient characteristics to determine which treatment option is most likely to result in a better outcome for the patient on average. Why is personalized medicine not done more in practice? One of many reasons is because practitioners do not have any easy way to holistically evaluate whether their personalization procedure does better than the standard of care. Our software, \qu{Personalized Treatment Evaluator} (the \proglang{R} package \pkg{PTE}), provides inference for improvement out-of-sample in many clinical scenarios. We also extend current methodology by allowing evaluation of improvement in the case where the endpoint is binary or survival. In the software, the practitioner inputs (1) data from a single-stage randomized trial with one continuous, incidence or survival endpoint and (2) a functional form of a model for the endpoint constructed from domain knowledge. The bootstrap is then employed on data unseen during model fitting to provide confidence intervals for the improvement for the average future patient (assuming future patients are similar to the patients in the trial). One may also test against a null scenario where the hypothesized personalization are not more useful than a standard of care. We demonstrate our method's promise on simulated data as well as on data from a randomized comparative trial investigating two treatments for depression.

% Personalized medicine involves the systematic use of individual patient characteristics to determine which treatment option is most likely to result in a better outcome for the patient on average. Why is it not done more in practice? One of many reasons is because practitioners do not have any easy way to holistically evaluate whether their personalization procedure does better than the standard of care. We provide a methodology with software (the R package PTE) that provides this evaluation by using statistical inference. We also extend current methodology by allowing this evaluation of improvement in the case where the endpoint is binary or survival. Our paper also provides a few literature review and a careful construction of the model, both of which are brushed over in most articles which do not seek to be didactic.

%\tiny
% \keyFont{ \section{Keywords:} personalized medicine, inference, bootstrap, treatment regime, randomized comparative trial, statistical software}
\end{abstract}

\section{Introduction}\label{sec:introduction}

Medical patients often respond differently to treatments and can experience varying side effects. Personalized medicine, sometimes called \qu{precision medicine} or \qu{stratified medicine} \citep{Smith2012}, is a medical paradigm offering the possibility for improving the health of individuals by judiciously treating individuals based on his or her heterogeneous prognostic or genomic information \citep{Zhao2013}. The interest in such personalization is exploding.

Fundamentally, personalized medicine is a statistical problem and much recent statistical research has focused on how to best estimate \textit{dynamic treatment regimes} or \textit{adaptive interventions} \citep{Collins2004, Chakraborty2014a}. These are essentially strategies that vary treatments administered over time as more is learned about how particular patients respond to one or more interventions. Elaborate models are often proposed that purport to estimate optimal dynamic treatment regimes from \textit{multi-stage} experiments \citep{Murphy2005a} as well as the more difficult situation of inference in observational studies.

The extant work, at least in the field of statistics, is highly theoretical. There is a dearth of software that can answer two fundamental questions practitioners will need answered before they can personalize future patients' treatments: (1) How much better is this personalization model expected to perform when compared to my previous \qu{naive} strategy for allocating treatments? (2) How confident can I be in this estimate? Can I reject a null hypothesis that it will perform no better than the standard of care? \citet[page 168]{Chakraborty2013} believe that \qu{more targeted research is warranted} on these important questions; and the goal of our paper is to provide a framework and usable software that fills in this gap. 

Personalized medicine is a broad paradigm encompassing many real-world situations. The setting we focus on herein is a common one: using previous randomized comparative / controlled trial (RCT) data to be able to make better decisions for future patients. We consider RCT's with two treatment options (two-arm), with one endpoint measure (also called the \qu{outcome} or \qu{response} which can be continuous, binary or survival) and where the researchers also collected a variety of patient characteristics to be used for personalization. The practitioner also has an idea of a model of the response (usually a simple first-order interaction model). Our software then answers the two critical questions listed above. 

Our advances are modest but important for practitioners. (1) Practitioners now have easy-to-use software that automates the testing of their personalization models. (2) We introduce a more intuitive metric that gauges how well the personalization is performing: \qu{improvement} versus a baseline strategy. (3) Our improvement estimates (and tests that they are statistically significantly better than nil) are all cross-validated, making the estimates honest if the data at hand was overfit and applicable to future patients. This external validity is only possible if future patients can be thought to come from the same population as the clinical trial, a generalization that is debated. (4) We have extended this well-established methodology for binary and survival endpoints, the most common endpoints in clinical trials.

The paper proceeds as follows. In Section~\ref{sec:background}, we review the modern personalized medicine literature and locate our method within. Section~\ref{sec:methodology} describes our methods and its limitations in depth, by describing the conceptual framework emphasizing our methodological advances. We then carefully specify the data and model inputs, define the improvement metric, and illustrate a strategy for providing practitioners with estimates and inference. Section~\ref{sec:data} applies our methods to (1) a simple simulated dataset in which the response model is known, (2) a more complicated dataset characterized by an unknown response model and (3) a real data set from a published clinical trial investigating two treatments for a major depressive disorder. Section~\ref{sec:software} demonstrates the software for all three types of endpoints: continuous, binary and survival. Section~\ref{sec:discussion} concludes and offers future directions of which there are many as our procedure is very limited.

\section{Background}\label{sec:background}

Consider an individual seeking one of two treatments, neither of which is known to be superior for all individuals. \qu{What treatment, by whom, is most effective for this individual with that specific problem, and under which set of circumstances?} \citep{Paul1967}.\footnote{Note that this problem is encountered in fields outside of just medicine. For example, finding the movie that will elicit the most enjoyment to the individual \citep{Zhou2008} or assessing wither a certain unepmployed individual be given job training \citep{LaLonde1986}. Although the methods discussed herein can be applied more generally, we will use examples and the vocabulary from the medical field.} Sometimes practitioners will select a treatment based informally on personal experience. Other times, practitioners may choose the treatment that their clinic or peers recommend. If the practitioner happens to be current on the research literature and there happens to be a published RCT whose results have clear clinical implications, the study's superior treatment (on average) may be chosen.

Each of these approaches can sometimes lead to improved outcomes, but each also can be badly flawed. For example, in a variety of clinical settings, \qu{craft lore} has been demonstrated to perform poorly, especially when compared to very simple statistical models \citep{Dawes1979}. It follows that each of these \qu{business-as-usual} \textit{treatment allocation procedures} can in principle be improved if there are patient characteristics available which are related to how well an intervention performs.% Patient \qu{characteristics} and \qu{circumstances}, also known as \qu{features,} \qu{states,} \qu{histories,} \qu{prognostic / prescriptive factors,} \qu{pretreatment variables,} and other terms of art, we will consider here to be \qu{covariates}, as they will be used in a regression modeling context.

%Should available patient covariates be related to the outcome of interest, one can imagine a model with inputs consisting of these covariates and the treatment. This model can provide a principled guess as to which treatment to administer compared to a business-as-usual approach. One can also imagine that estimates of \qu{how much better} could be computed. The preferred treatment could then be individually selected for each patient.

The need for personalized medicine by using patient characteristics is by no means a novel idea. As noted as early as 1865, \qu{the response of the average patient to therapy is not necessarily the response of the patient being treated} (translated by \citealt{Bernard1957}). There is now a substantial literature addressing numerous aspects of personalized medicine. The field is quite fragmented: there is literature on treatment-covariate interactions, locating subgroups of patients and personalized treatment effects estimation. A focus on inference is rare in the literature and available software for inference is negligible.

\citet{Byar1985} provides an early review of work involving treatment-covariate interactions. \citet{Byar1977} investigates tests for treatment-covariate interactions in survival models and discusses methods for treatment recommendations based on covariate patterns. \citet{Shuster1983} considers two treatments and proposes a linear model composed of a treatment effect, a prognostic factor, and their interaction. Using this model, the authors create confidence intervals to determine for which values of the prognostic factor one of two treatments is superior. 

Many researchers also became interested in discovering \qu{qualitative interactions}, which are interactions that create a subset of patients for which one treatment is superior and another subset for which the alternative treatment is superior. \citet{Gail1985} develop a likelihood ratio test for qualitative interactions which was further extended by \citet{Pan1997} and \citet{Silvapulle2001}. For information and another approach, see \citet{Foster2013}.

Much of the early work in detecting these interactions required a prior specification of subgroups. This can present significant difficulties in the presence of high dimensionality or complicated associations. More recent approaches such as \citet{Su2009} and \citet{Dusseldorp2014} favor recursive partitioning trees that discover important nonlinearities and interactions. \citet{Dusseldorp2016} introduce an \proglang{R} package called \pkg{QUINT} that outputs binary trees breaking participants into subgroups. \citet{Shen2016} propose a kernel machine score test to identify interactions and the test has more power than the classic Wald test when the predictor effects are non-linear and when there is a large number of predictors. \citet{Berger2014} discuss a method for creating prior subgroup probabilities and provides a Bayesian method for uncovering interactions and identifying subgroups.

In our method, we make use of RCT data. Thus, it is important to remember that \qu{clinical trials are typically not powered to examine subgroup effects or interaction effects, which are closely related to personalization... even if an optimal personalized medicine rule can provide substantial gains it may be difficult to estimate this rule with few subjects} \citep{Rubin2012}. This is why a major bulk of the literature focuses on not finding covariate-treatment interactions or locating subgroups of individuals, but the entire model itself, \qu{regimes}, that is then used to sort individuals. Holistic statements can then be made on the basis of this entire sorting procedure. We turn to some of this literature now.

\citet{Zhang2012a} consider the context of treatment regime estimation in the presence of model misspecification when there is a single-point treatment decision. By applying a doubly-robust augmented inverse probability weighted estimator that under the right circumstances can adjust for confounding and by considering a restricted set of policies, their approach can help protect against misspecification of either the propensity score model or the regression model for patient outcome. \citet{Brinkley2010} develop a regression-based framework of a dichotomous response for personalized treatment regimes within the rubric of \qu{attributable risk}. They propose developing optimal treatment regimes that minimize the probability of a poor outcome, and then consider the positive consequences, or \qu{attributable benefit}, of their regime.  They also develop asymptotically valid inference for a parameter similar to improvement with business-as-usual as the random (see our Section~\ref{naive_allocation_procedures}), an idea we extend. Within the literature, their work is the closest conceptually to ours. \citet{Gunter2011} develop a stepwise approach to variable selection and in \citet{Gunter2011a} compare it to stepwise regression. Rather than using a traditional sum-of-squares metric, the authors' method compares the estimated mean response, or \qu{value,} of the optimal policy for the models considered, a concept we make use of in Section~\ref{sec:methodology}. \citet{Imai2013} use a modified Support Vector Machine with LASSO constraints to select the variables useful in an optimal regime when the response is binary. \citet{Laan2013} uses a loss-based super-learning approach with cross-validation.

Also important within the area of treatment regime estimation, but not explored in this paper, is the estimation of \textit{dynamic treatment regimes} (DTRs). DTRs constitute a set of decision rules, estimated from many experimental and longitudinal intervals. Each regime is intended to produce the highest mean response over that time interval. Naturally, the focus is on \textit{optimal} DTRs --- the decision rules which provide the highest mean response. \citet{Murphy2003} and \citet{Robins2004} develop two influential approaches based on regret functions and nested mean models respectively. \citet{Moodie2007} discuss the relationship between the two while \citet{Moodie2009} and \citet{Chakraborty2009} present approaches for mitigating biases (the latter also fixes biases in model parameter estimation stemming from their non-regularity in SMART trials). \citet{Robins2008} focus on using observational data and optimizing the time for administering the stages --- the \qu{when to start} --- within the DTR. \citet{Orellana2010} develop a different approach for estimating optimal DTRs based on marginal structural mean models. \citet{Henderson2010} develop optimal DTR estimation using regret functions and also focus on diagnostics and model checking. \citet{Barrett2013} develop a doubly robust extension of this approach for use in observational data. \citet{Laber2014} demonstrate the application of set-valued DTRs that allow balancing of multiple possible outcomes, such as relieving symptoms or minimizing patient side effects. Their approach produces a subset of recommended treatments rather than a single treatment. Also, \citet{McKeague2014} estimate treatment regimes from functional predictors in RCTs to incorporate biosignatures such as brain scans or mass spectrometry.

Many of the procedures developed for estimating DTRs have roots in reinforcement learning. Two widely-used methods are Q-learning \citet{Murphy2005b} and A-learning (see \citealt{Schulte2012} for an overview of these concepts). One well-noted difficulty with Q-learning and A-learning are their susceptibility to model misspecification. Consequently, researchers have begun to focus on \qu{robust} methods for DTR estimation. \citet{Zhang2013} extend the doubly-robust augmented inverse probability weighted method \citep{Zhang2012a} by considering multiple binary treatment stages.

Many of the methods mentioned above can be extended to censored survival data. \citet{Zhao2015} describe a computationally efficient method for estimating a treatment regimes that maximizes mean survival time by extending the weighted learning inverse probability method. This method is doubly robust; it is protected from model misspecification if either the censoring model or the survival model is correct. Additionally, methods for DTR estimation can be extended. \citet{Goldberg2012} extend Q-learning  with inverse-probability-of-censoring weighting to find the optimal treatment plan for individual patients, and the method allows for flexibility in the number of treatment stages. 

It has been tempting, when creating these treatment regime models, to directly employ then to predict the differential response of individuals among different treatments. This is called in the literature \qu{heterogeneous treatment effects models} or \qu{individualized treatment rules} and there is quite a lot of interest in it. 

Surprisingly, methods designed for accurate estimation of an overall conditional mean of the response may not perform well when the goal is to estimate these individualized treatment rules. \citet{Qian2011} propose a two-step approach to estimating \qu{individualized treatment rules} based on single-stage randomized trials using $\ell_1$-penalized regression while \citet{Lu2011, Lu2013} use quadratic loss which facilitates variable selection. \citet{Rolling2014} develop a new form of cross-validation which chooses between different heterogeneous treatment models.

One current area of research in heterogeneous effect estimation is the development of algorithms that can be used to create finer and more accurate partitions. \citet{Kallus2017} presents three methods for the case of observational data: greedily partitioning data to find optimal trees, bootstrap aggregating to create a \qu{personalization forest} a la Random Forests, and using the tree method coupled with mixed integer programming to find the optimal tree. \citet{Lamont2016} build on the prior methods, parametric multiple imputation and recursive partitioning, to estimate heterogeneous treatment effects and compare the performance of both methods. This estimation can be extended to censored data. \citet{Henderson2017} discuss the implementation of Bayesian additive regression trees for estimating heterogeneous effects, and they can be used for continuous, binary and censored data. \citet{Ma2019} proposes a Bayesian predictive method that integrates multiple sources of biomarkers.

One major drawback of many of the approaches in the literature reviewed is their significant difficulty evaluating estimator performance. Put another way, given the complexity of the estimation procedures, statistical inference is very challenging. Many of the approaches require that the proposed model be correct. There are numerous applications in the biomedical sciences for which this assumption is neither credible nor testable in practice. For example, \citet{Evans2004} consider pharmacogenomics, and argue that as our understanding of the genetic influences on individual variation in drug response and side-effects improves, there will be increased opportunity to incorporate genetic moderators to enhance personalized treatment. But we will ever truly understand this model well enough to properly specify it? Further, other biomarkers (e.g. neuroimaging) of treatment response have begun to emerge, and the integration of these diverse moderators will require flexible approaches that are robust to model misspecification \citep{McGrath2013}. How will the models of today incorporate important relationships that can be anticipated but have yet to be identified? Further, many proposed methods employ non-parametric models use the data to decide which internal parameters to fit and then in turn estimates these internal parameters. Thus a form of model selection that introduces difficult inferential complications (see \citealp{Berk2013b}).

At the very least, therefore, there should be an alternative inferential framework for evaluating treatment regimes that do not require correct model specification (and thereby obviating the need for model checking and diagnostics) nor knowledge of unmeasured characteristics (see discussion in \citealp{Henderson2010}) accompanied by easy-to-use software. This is the modest goal herein.

\section{Methods}\label{sec:methodology}

This work seeks to be didactic and thus carefully explains the extant methodology and framework (Section~\ref{subsec:conceptual_framework}), data inputs (Section~\ref{subsec:required_inputs}), estimation (Section~\ref{subsec:I_scores}) and a procedure for inference (Section~\ref{subsec:cis_tests_I_measure}).  For those familiar with this literature, these sections can be skipped. Our methodological contributions then are to (a) employ out-of-sample validation to  (Section~\ref{subsec:I_scores}) specifically to the (b) \emph{improvement}, the metric defined as the difference in between how patients allocated via personalized model fare in their outcome and a patients allocated via business-as-usual model fare in their outcome (Section~\ref{subsec:naive_allocation_procedures}) and (c) to extend this validation methodology to binary and survival endpoints (Sections~\ref{subsec:I_scores_binary} and \ref{subsec:I_scores_survival}). Table~\ref{tab:notation} serves as a guide to the main notation used in this section.\\

\begin{table}[p]
\centering
\caption{A compendium of the main notation in our methodology by section.}
\begin{tabular}{ll}
Notation & Description \\ \hline
\multicolumn{2}{l}{Framework (Section~\ref{subsec:conceptual_framework})} \\ \hline
$Y$ & The random variable (r.v) for the outcomes for the subjects. \\
$\X$, $\mathcal{X}$ & The r.v. for the observed measurements for the subjects, its support. \\
$A$ & The r.v. for the treatment. \\
$T_1, T_2$ & The two treatments, shorthand for their codes, zero and one. \\ 
$d$ or $d[f]$ & The decision function; it maps observed measurements to treatment.\\
$V$ or $V[d]$ & The value of the decision function; it is the average outcome over all  \\
& patients if this decision is used to allocate treatment.\\
$d^*$ & The unknown optimal decision function i.e. the one with highest $V$.\\
$d_0$ & A naive, baseline, business-as-usual or null decision function.\\
$\muIo$ & The unknown improvement of $d$ over $d_0$; it is the difference of values. \\
\hline
\multicolumn{2}{l}{The RCT data (Section~\ref{eq:rct_data_notation})} \\ \hline
$n$ & The number of subjects in the randomized comparative trial (RCT). \\
$p$ & The number of measurements taken on each subject. \\
$\x_i$ & The vector of $p$ measurements for the $i$th subject. \\
$x_{i,j}$ & The $j$th measurement for the $i$th subject. \\
$\X$ & The $n \times p$ matrix of all measurements for all subjects.\\
$\A$ & The vector of treatments for all the $n$ subjects. \\
$\y$ & The vector of outcomes for all the $n$ subjects. \\ \hline
\multicolumn{2}{l}{The response model (Section~\ref{subsec:model_f})} \\ \hline
$f$ & The function that relates the $p$ measurements and $A$ to the response. \\
%$\thetavec$ & Fixed parameters needed to compute $f$.\\
$U_i$ & The r.v. for the unknown covariates for subject $i$.\\
$\xi(\x_i, A_i, U_i)$ & The function that computes misspecification in the response.\\
$\errorrv_i$ & The r.v. for irreducible noise for the $i$th subject.\\
$\beta_j$ & The linear coefficient for the $j$th measurement when $A = 0$. \\
$\gamma_j$ & The additional linear coefficient for the $j$th measurement when $A = 1$. \\ \hline
\multicolumn{2}{l}{Out of sample estimation and validation (Section~\ref{subsec:I_scores})} \\ \hline
$\Xtrain, \ytrain$ & The subset of the data used to create the fit of $f$.\\
$\Xtest, \ytest$ & The subset of the data used to validate the fit of $f$.\\
$\hat{f}$, $\hat{d}$, $\hat{V}$, $\hat{I}_0$ & The finite-sample estimates of $f$, $d$, $V$, $\muIo$. \\
$\ybar_{set}$ & The arithmetic average of $\braces{y_i~:~i \in set}$. \\
$\hat{\beta}_j$, $\hat{\gamma}_j$ & The finite-sample estimates of $\beta_j$, $\gamma_j$. \\ \hline
\multicolumn{2}{l}{Inference (Section~\ref{subsec:cis_tests_I_measure})} \\ \hline
$B$ & The number of bootstrap samples. \\
$\tilde{\X}, \tilde{\y}$ & A sample of the rows of $\X, \y$ with replacement.\\
$\tilde{I}_{0, b}$ & The $b$th estimate of $\muIo$ in the bootstrap. \\
$\alpha$ & The size of the hypothesis test.\\ \hline
\multicolumn{2}{l}{Personalization of future subjects' treatments (Section~\ref{subsec:future_subjects})} \\ \hline
$\x_*$ & A future subject (not part of the RCT). \\
%$$ & \\
%$$ & \\
%$$ & \\
%$$ & \\
\end{tabular}
\label{tab:notation}
\end{table}

\subsection{Conceptual Framework}\label{subsec:conceptual_framework}

We imagine a set of random variables having a joint probability distribution that can be properly seen as a population from which data could be randomly and independently realized. The population can also be imagined as all potential observations that could be realized from the joint probability distribution. Either conception is consistent with our setup. 

A researcher chooses one of the random variables to be the response $Y$ which could be continuous, binary or survival (with a corresponding censoring variable, explained later). We assume without loss of generality that a greater-valued outcome is better for all individuals. Then, one or more of the other random variables are covariates $\X \in \mathcal{X}$. At the moment, we do not distinguish between observed and unobserved covariates but we will later. There is then a conditional distribution $\cprob{Y}{\X}$ whose conditional expectation $\cexpe{Y}{\X}$ constitutes the overall population response surface. No functional forms are imposed and for generality we allow the functional form to be nonlinear with interactions among the covariates which is the case in artificial intelligence procedures (machine learning).

All potential observations are hypothetical study subjects. Each can be exposed to a random treatment denoted $A \in \braces{0, 1}$ where zero codes for the first experimental condition also equivalently referred to as $T_1$ (which may be considered the \qu{control} or \qu{comparison} condition) and one codes for another experimental condition equivalently referred to as $T_2$. We make the standard assumption of no interference between study subjects, which means that the outcome for any given subject is unaffected by the interventions to which other subjects are randomly assigned \citep{Cox1958} and outcomes under either condition can vary over subjects \citep[Section 2.5.1]{Rosenbaum2002}. In short, we employ the conventional Neyman-Rubin approach \citep{Rubin1974} but treat all the data as randomly realized \citep{Berk2013}. 

A standard estimation target in RCTs is the \textit{population average treatment effect} (PATE), defined here as $\cexpe{Y}{A = 1} - \cexpe{Y}{A = 0}$, the difference between the population expectations. That is, the PATE is defined as the difference in mean outcome were all subjects exposed to $T_{2}$ or alternatively were all exposed to $T_{1}$. In a randomized controlled trial, the PATE is synonymous with the overall efficacy of the treatment of interest and it is almost invariably the goal of the trial \citep{Zhao2013}.

For personalization, we want to make use of any association between $Y$ and $\X$. For the hypothetical study subjects, there is a conditional population response surface $\cexpe{Y}{\X, A = 1}$ and another conditional population response surface $\cexpe{Y}{\X, A = 0}$, a key objective being to exploit the difference in these response surfaces for better treatment allocation. The typical approach is to create a deterministic \textit{individualized treatment decision rule} $d$ that takes an individual's covariates and maps them to a treatment. We seek $d : \mathcal{X} \rightarrow \braces{0, 1}$ based on knowledge of the differing conditional population response surfaces. The rule is sometimes called an \textit{allocation procedure} because it determines which treatment to allocate based on measurements made on the individual. To compare different allocation procedures, our metric is the expectation of the outcome $Y$ using the allocation procedure $d$ averaged over all subjects $\mathcal{X}$. Following the notation of \citet{Qian2011}, we denote this expectation as the \textit{value} of the decision rule
% and using Equation~\ref{eq:joint_dist}, we obtain

\beqn%\label{eq:def_of_value}
V[d] := \expesubsup{\X,A}{d}{Y} \triangleq \int_{\mathcal{X}} \parens{\sum_{a \in \braces{0, 1}} \parens{ \int_{\reals} y f_{Y | \X, A}(y, \x, a) dy} \indic{a = d(\x)} } f_{\X}(\x) d\x. 
\eeqn

\noindent Although the integral expression appears complicated, when unpacked it is merely an expectation of the response averaged over $\mathcal{X}$, the space of all patients characteristics. When averaging over $\mathcal{X}$, different treatments will be recommended based on the rule, i.e. $a = d(\x)$, and that in turn will modify the density of the response, $f_{Y | \X}$. Put another way, $V[d]$ is the mean patient outcome when personalizing each patient's treatment.

We have considered all covariates to be random variables because we envision \textit{future} patients for whom an appropriate treatment is required. Ideally, their covariate values are realized from the same joint distribution as the covariate values for the study subjects, an assumption that is debated and discussed in the concluding section. 

In addition, we do not intend to rely on estimates of the two population response surfaces. As a practical matter, we will make do with a \textit{population} response surface \emph{approximation} for each. No assumptions are made about the nature of these approximations and in particular, how well or poorly either population approximation corresponds to the true conditional response surfaces. 

Recall that much of the recent literature has been focused on finding the \textit{optimal} rule, $d^* \triangleq \argmax_{d} \braces{V[d]}$. Although this is an admirable ideal (as in \citealt{Qian2011}), our goals here are more modest. We envision an imperfect rule $d$ far from $d^*$, and we wish to gauge its performance relative to the performance of another rule $d_0$, where the \qu{naught} denotes a business-as-usual allocation procedure, sometimes called \qu{standard of care}. Thus, we define the population value \textit{improvement} $\muIo$ as the value of $d$ minus the value of $d_0$,

\bneqn\label{eq:mean_improvement_score}
\muIo \triangleq V[d]- V[d_0] = \expesubsup{\X,A}{d}{Y} - \expesubsup{\X,A}{d_0}{Y}
\eneqn

which is sometimes called \emph{benefit} in the literature. Since our convention is that higher response values are better, we seek large, positive improvements that translate to better average performance (as measured by the response). Note that this is a natural measure when $Y$ is continuous. When $Y$ is incidence or survival, we redefine $\muIo$ (see Sections~\ref{subsec:I_scores_binary} and \ref{subsec:I_scores_survival}).

The metric $\muIo$ is not standard in the literature but we strongly believe it to be the natural metric for personalization following \citet{Kallus2017}. There are many other such metrics beyond value and improvement. For example, \citet{Ma2019} uses three (1) the expected number of subjects misassigned to their optimal treatment, (2) exepected gain or loss in treatment utility and (3) the expected proportion where the model correctly predicted the response which is useful only in the binary response case which we address later. \\

\subsection{Our framework's required inputs}\label{subsec:required_inputs}

Our method depends on two inputs (1) access to RCT data and (2) either a prespecified parametric model $f(\x, A, \thetavec)$ for the population approximation of the true response surfaces or an explicit $d$ function. If we prespecified $f$, we then use the RCT data to estimate parameters of the model $\hat{\thetavec}$, and the estimates are embedded in the estimated model, $\fhat$. This model estimate permits us, in turn, to construct an estimated decision rule $\dhat$ and an estimate of the improved outcomes future subjects will experience (explained later in Section~\ref{subsec:I_scores}). We assume that the model $f$ is specified before looking at the data. \qu{Data snooping} (running our method, checking the $p$-value, changing the model $f$ and running again) fosters overfitting and can introduce serious estimation bias, invalidating confidence intervals and statistical tests \citep{Berk2013b}.\\

\subsubsection{The RCT data}\label{eq:rct_data_notation}

Our procedure strictly requires RCT data to ensure there is a causal effect of the heterogeneous parameters. Much of the research discussed in the background (Section~\ref{sec:background}) applies in the case of observational data. We realize this limits the scope of our proposal. The RCT data must come from an experiment undertaken to estimate the PATE for treatments $T_1$ and $T_2$ for a diagnosis of a disease of interest. $T_1$ and $T_2$ are the same treatments one would offer to future subjects with the same diagnosis. %We believe our method can work with any randomization (allocation) scheme widely used in practice today (these allocations do not unfairly allocate to one treatment over another in regions of covariate space). 

There are $n$ subjects each with $p$ covariates which are denoted for the $i$th subject as $\x_i \triangleq \bracks{x_{i1}, x_{i2}, \ldots, x_{ip}}$. Because these covariates will be used to construct a decision rule applied with future patients in clinical settings, the $\x_i$'s in the RCT data must be the same covariates measured for new subjects. Thus, such characteristics such as the site of treatment (in a multi-center trial) or the identification of the medical practitioner who treated each subject or hindsight-only variables are not included.

We assume the outcome measure of interest $y_i$ is assessed once per subject. Aggregating all covariate vectors, binary allocations and responses row-wise, we denote the full RCT data as the column-bound matrix $\bracks{\X, \A, \y}$.  In practice, missing data can be imputed (in both the RCT data and the future data), but herein we assume complete data. 

We will be drawing inference to a patient population beyond those who participated in the experiment. Formally, new subjects must be sampled from that same population as were the subjects in the RCT. In the absence of explicit probability sampling, the case would need to be made that the model can generalize. This requires subject-matter expertise and knowledge of how the study subjects were recruited.\\

\subsubsection{The Model for the Response based on Observed Measurements}\label{subsec:model_f}

The decision rule $d$ is a function of $\x$ through $f$ and is defined as

\bneqn\label{eq:d}
d[f(\x)] \triangleq \argmax_{A \in \braces{0, 1}} f(\x, A) = \indic{f(\x, 1) - f(\x, 0)}.
\eneqn

\noindent As in \citet{Berk2014}, we assume the model $f$ provided by the practitioner to be an \textit{approximation using the available information},

\bneqn\label{eq:response_model_decomposition}
Y_i = \underbrace{f(\X_i, A_i) + \xi(\X_i, \U_i, A_i)}_{\cexpe{Y_i}{\X_i, \U_i, A_i}} + \errorrv_i,
\eneqn

\noindent where $f$ differs from the true response expectation by a term dependent on $\U$, the unobserved information. The last term $\errorrv_i$ is the irreducible noise around the true conditional expectations and is taken to be independent and identically distributed, mean-centered and uncorrelated with the covariates. Even in the absence of $\errorrv_i$, $f$ will always differ from the true conditional expectation function by $\xi_i(\X_i, \U_i, A_i)$, which represents model misspecification \citep[Chapter 13]{Box1987}.

We wish only to determine whether an estimate of $f$ is \textit{useful} for improving treatment allocation for future patients (that are similar to the patients in the RCT) and do not expect to recover the optimal allocation rule $d^*$ which requires the unseen $\U$. Further, we do not concern ourselves with substantive interpretations associated with any of the $p$ covariates, a goal of future research. Thus, our method is robust to model misspecification by construction. %One implication is that a wide variety of models and estimation procedures for $f$ could in principle prove useful. 

What could $f$ look like in practice? Assume a continuous response (binary and survival are discussed later) and consider the conventional linear regression model with first order interactions. Much of the literature we reviewed in Section~\ref{sec:background} favored this class of models. We specify a linear model containing a subset of the covariates used as main effects and a possibly differing subset of the covariates to be employed as first order interactions with the treatment indicator, $\braces{x_{1'}, \ldots, x_{p'}} \subset \braces{x_1, \ldots, x_p}$, selected using domain knowledge:

\bneqn\label{eq:provided_model}
f(\x_{i1}, A_i) = \beta_0 + \beta_{1}x_{1} + \ldots + \beta_{p}x_{p} + A_i \parens{\gamma_{0} + \gamma_{1'} x_{1'} + \ldots + \gamma_{p'}x_{p'}}.
\eneqn

\noindent These interactions induce heterogeneous effects between $T_1$ and $T_2$ for a subject $\x$ in a very interpretable way: $d[f(\x)] = 1$ when $\gamma_{0} + \gamma_{1'} x_{1'} + \ldots + \gamma_{p'}x_{p'} > 0$ and 0 otherwise. The $\gamma$'s are the critical component of the model if there are systematic patient-specific responses to the interventions. Thereby, $d$ varies over different points in $\mathcal{X}$ space. Note that rules derived from this type of conventional model also have the added bonus as being interpretable as a best linear approximation of the true relationship.

We stress that our models are \textit{not} required to be of this form, but we introduce them here mostly for familiarity and pedagogical simplicity. There are times when these models will perform terribly even if $\braces{x_{1'}, \ldots, x_{p'}}$ are the correct moderating variables. For a non-linear example, see \citet[Fig 1, right]{Zhao2013}. Although this model is the default implementation, the user can specify any model desired in the software. This will be discussed in Section~\ref{sec:software}.

We also stress that although the theory for estimating linear models' coefficients (as well as those for logistic regression and Weibull regression) is well-developed, we are not interested in inference for these coefficients in this work as our goal is only estimation and inference for overall usefulness of the personalization scheme, i.e. the unknown parameter $\muIo$. This will become clear in the next few sections.\\

\subsection{Other allocation procedures}\label{subsec:naive_allocation_procedures}

Although $d_0$ can be any allocation rule, for the purposes of the paper, we examine only two ``business-as-usual'' allocation procedures (others are discussed as extensions in Section~\ref{sec:discussion}). The first we call \textbf{random} denoting the allocation where the patient receives $T_1$ or $T_2$ with a fair coin flip, probability 50\%. This serves as a baseline or \qu{straw man} but nevertheless an important standard --- the personalization model should be able to provide better patient outcomes than a completely random allocation.

The second business-as-usual procedure we call \textbf{best}. This procedure gives all patients the better of the two treatments as determined by the comparison of the sample average for all subjects who received $T_1$ denoted $\ybar_{T_1}$ and the sample average of all subjects who received $T_2$ denoted $\ybar_{T_2}$. This is used as the default in many frameworks for example \citet{Kang2014}. We consider beating this procedure the gold standard in proof that the personalization truly works as as practitioners most often employ the current best known treatment. However, some consider this comparison conservative \citep[Section 7]{Brinkley2010} and the next section will describe why it is statistically conservative as we lose sample size when demanding this comparison. Due to this conservativeness, barring conclusive evidence that either $T_1$ or $T_2$ is superior, the \textbf{random} procedure should be the standard of comparison. This case is not infrequent in RCTs which feature negative comparison results, the case in our clinical trial example of Section~\ref{subsec:clinical_trial_example}. \\

\subsection{Estimating the improvement scores}\label{subsec:I_scores}

\subsubsection{For a Continuous Response}\label{subsec:I_scores_continuous}

How well do unseen subjects with treatments allocated by $d$ do on average compared to the same unseen subjects with treatments allocated by $d_0$? We start by computing the \textit{estimated improvement score}, a sample statistic given by

\bneqn\label{eq:estimated_improvement_score}
\Ihato \triangleq \Vhat[\dhat] - \Vhat[\hat{d_0}],
\eneqn

\noindent where $\dhat$ is an estimate of the rule $d$ derived from the population response surface approximation, $\Vhat$ is an estimate of its corresponding value $V$ and $\Ihato$ is an estimate of the resulting population improvement $\muIo$ (Equation~\ref{eq:mean_improvement_score}). The $\hat{d_0}$ notation indicates that sometimes the competitor $d_0$ may have to be estimated from the data as well. For example, the allocation procedure \textbf{best} must be calculated by using the sample average of the responses for both $T_1$ and $T_2$ in the data.

In order to properly estimate $\muIo$, we use the widely-known cross-validation procedure \citep[Chapter 7.10]{Hastie2013}. We split the RCT data into two disjoint subsets: training data with $\ntrain$ of the original $n$ observations $\bracks{\Xtrain, \ytrain}$ and testing data with the remaining $\ntest = n - \ntrain$ observations $\bracks{\Xtest, \ytest}$. Then $\fhattrain$ can be fit using the training data to construct $\dhat$ via Equation~\ref{eq:d}. Performance of $\dhat$ as calculated by Equation~\ref{eq:estimated_improvement_score}, is then evaluated on the test data. \citet{Hastie2013} explain that a single train-test split yields an estimate of the \qu{performance} of the procedure on future individuals conditional on $\bracks{\Xtrain, \ytrain}$, the \qu{past}. Thus, the $\Ihato$ statistic defined in Equation~\ref{eq:estimated_improvement_score} computed using $\bracks{\Xtest, \ytest}$ can provide an honest assessment of improvement (i.e. immune to overfitting in $\fhat$) who are allocated using our proposed methodology compared to a baseline business-as-usual allocation strategy \citep{Faraway2013}. This can be thought of as employing a replicated trial, often required in drug development programs, which separates rule construction (in-sample) from rule validation (out-of-sample) as recommended by \citet{Rubin2012}. Note that this comes at a cost of more sample variability (as now our estimate will be based on the test subset with a sample size much smaller than $n$). Our framework and software is the first to provide user-friendly out-of-sample validation for the overall utility of personalized medicine models as a native feature. 

Given the estimates $\hat{d}$ and $\hat{d_0}$, the question remains of how to explicitly compute $\Vhat$ for subjects we have not yet seen in order to estimate $\Ihato$. That is, we are trying to estimate the expectation of an allocation procedure over covariate space $\mathcal{X}$. 

Recall that in the test data, our allocation prediction $\dhat(\x_i)$ is the binary recommendation of $T_1$ or $T_2$ for each $\xtesti$. If we recommended the treatment that the subject actually was allocated in the RCT, i.e. $\dhat(\x_i) = A_i$, we consider that subject to be \qu{lucky}. We define lucky in the sense that by the flip of the coin, the subject was randomly allocated to the treatment that our model-based allocation procedure estimates to be the better of the two treatments. 

The average of the lucky subjects' responses should estimate the average of the response of new subjects who are allocated to their treatments based on our procedure $d$ and this is the estimate of $\Vhat[\dhat]$ we are seeking. Because the $\x$'s in the test data are assumed to be sampled randomly from population covariates, this sample average estimates the expectation over $\mathcal{X}$, i.e. $\expesubsup{\X, A}{d}{Y}$ conditional on the training set. In order to make this concept more clear, it is convenient to consider Table~\ref{tab:two_by_two}, a $2 \times 2$ matrix which houses the sorted entries of the out-of-sample $\ytest$ based on the predictions, the $\dhat(\x_i)$'s. 

\begin{table}[h]
\centering
\begin{tabular}{r|c|c}
& $\dhat(\x_i) = 0$ & $\dhat(\x_i) = 1$ \\\hline
$A_i = 0$ & $P$ & $Q$ \\\hline
$A_i = 1$ & $R$ & $S$ \\
\end{tabular}
\caption{The elements of $\ytest$ cross-tabulated by their administered treatment $A_i$ and our model's estimate of the better treatment $\dhat(\x_i)$.}
\label{tab:two_by_two}
\end{table}

The diagonal entries of sets $P$ and $S$ contain the \qu{lucky} subjects. The notation $\ybar_{\cdot}$ indicates the sample average among the elements of $\ytest$ specified in the subscript located in the cells of the table.

How do we compute  $\Vhat[\dhat_0]$, the business-as-usual procedure? For \textbf{rand}, we simply average all of the $\ytest$ responses;  for \textbf{best}, we average the $\ytest$ responses for the treatment group that has a larger sample average. Thus, the sample statistics of Equation~\ref{eq:estimated_improvement_score} can be written as

\bneqn\label{eq:I_random}
\hat{I}_{\text{random}} &\triangleq& \ybar_{P \cup S} - \ybar_{\text{test}}, \label{eq:I_rand}\\
\hat{I}_{\text{best}} &\triangleq& \ybar_{P \cup S} - \begin{cases}
\ybar_{P \cup Q} \quad\text{when}\quad \ybar_{P \cup Q} \geq \ybar_{R \cup S} \\
\ybar_{R \cup S} \quad\text{when}\quad \ybar_{P \cup Q} < \ybar_{R \cup S} \label{eq:I_best}.\\
\end{cases}
\eneqn

\noindent Note that the plug-in estimate of value $\Vhat[\dhat] = \ybar_{P \cup S}$ is traditional in the personalized medicine literature. For example, in \citet[Corollary 3]{Kallus2017} it is written as

\bneqn\label{value_estimate_non_smooth}
\Vhat[\dhat] := \sum_{i=1}^n Y_i \indic{\dhat(\x_i) =A_i} \Biggm/  \sum_{i=1}^n \indic{\dhat(\x_i) =A_i}.
\eneqn

There is one more conceptual point. Recall that the value estimates $\Vhat\bracks{\cdot}$ are conditional on the training set. This means they do not estimate the unconditional $\expesubsup{\X,A}{d}{Y}$. To address this, \citet[Chapter 7]{Hastie2013} recommend that the same procedure be performed across many different mutually exclusive and collectively exhaustive splits of the full date. This procedure of building many models is called \qu{$K$-fold cross-validation} (CV) and its purpose is to integrate out the effect of a single training set to result in the unconditional estimate of generalization. This is \qu{an alternative approach ... [that] for simplicity ... [was not] consider[ed] ... further} in the previous investigation of \citet[page 5]{Chakraborty2014b}.

In practice, how large should the training and test splits be? Depending on the size of the test set relative to the training set, CV can trade bias for variance when estimating an out-of-sample metric. Small training sets and large test sets give more biased estimates since the training set is built with less data than the $n$ observations given. However, large test sets have lower variance estimates since they are composed of many examples. There is no consensus in the literature about the optimal training-test split size \citep[page 242]{Hastie2013} but 10-fold CV is a common choice employed in many statistical applications and provides for a relatively fast algorithm. In the limit, $n$ models can be created by leaving each observation out, as done in \citet{DeRubeis2014}. In our software, we default to 10-fold cross validation but allow for user customization. 

\begin{figure}[htp]
\centering
\includegraphics[width=5.5in]{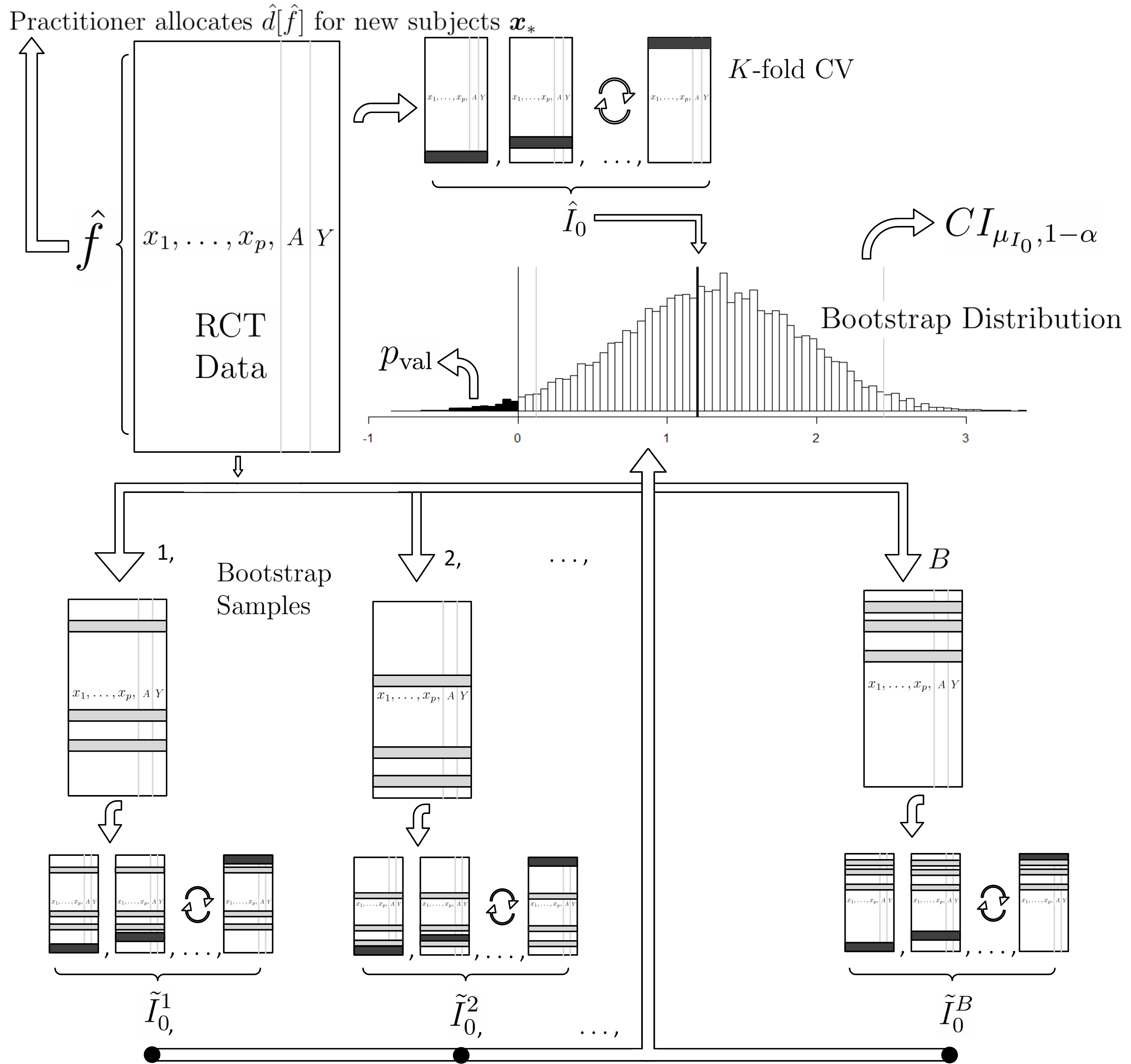}
\caption{A graphical illustration of (1) our proposed method for estimation and (2) our proposed method for inference on the population mean improvement of an allocation procedure and (3) our proposed future allocation procedure. To compute the best estimate of the improvement $\Ihato$, the RCT data goes through the $K$-fold cross validation procedure of Section~\ref{subsec:I_scores} (depicted in the top center). The black slices of the data frame represent the test data. To draw inference, we employ the non-parametric bootstrap procedure of Section~\ref{subsec:cis_tests_I_measure} by sampling the RCT data with replacement and repeating the $K$-fold CV to produce $\Ihato^1,~\Ihato^2, \ldots, \Ihato^B$ (bottom). The grey slices of the data frame represent the duplicate rows in the original data due to sampling with replacement. The confidence interval and significance of $H_0: \muIo \leq 0$ is computed from the bootstrap distribution (middle center). Finally, the practitioner receives $\fhat$ which is built with the complete RCT data (top left).}
\label{fig:method_illustration}
\end{figure}

This estimation procedure outlined above is graphically illustrated in the top of Fig~\ref{fig:method_illustration}. We now extend this methodology to binary and survival endpoints in the next two sections.

\subsubsection{For a Binary Response}\label{subsec:I_scores_binary}

In the binary case, we let $V$ be the expected probability of the positive outcome under $d$ just like in \citet{Kang2014}. There are three choices of improvement metrics (a) the probability difference, (b) the risk ratio and (c) the odds ratio:

\beqn%\label{eq:estimated_improvement_score_binary}
\text{(a)} \quad \muIo \triangleq V[d] - V[{d_0}], \quad\quad
\text{(b)} \quad \muIo \triangleq \frac{V[d]}{V[{d_0}]} \quad\quad
\text{and (c)} \quad \muIo \triangleq \frac{
\frac{V[d]}{1 - V[d]}
}{
\frac{V[{d_0}]}{1 - V[{d_0}]}.
}
\eeqn

\noindent and the estimate of all three (the probability difference, the risk ratio and the odds ratio) is found by placing hats on each term in the definitions above (all $V$'s, $d$'s and $d_0$'s). 

Following the example in the previous section we employ the analogous model, a logistic linear model with first order treatment interactions where the model $f$ now denotes the probability of the positive outcome $y=1$,

\bneqn\label{eq:logit_model}
f(\x_{i1}, A_i) = \text{logit}\parens{\beta_0 + \beta_{1}x_{1} + \ldots + \beta_{p}x_{p} + A_i \parens{\gamma_{0} + \gamma_{1'} x_{1'} + \ldots + \gamma_{p'}x_{p'}}}.
\eneqn

\noindent This model, fit via maximum likelihood numerically \citep{Agresti2013}, is the default in our software implementation. Here, higher probabilities of success imply higher logit values so that we have the same form of the decision rule estimate, $\hat{d}[\fhat(\x)] = 1$ when $\hat{\gamma}_{0} + \hat{\gamma}_{1'} x_{1'} + \ldots$ $+\hat{\gamma}_{p'}x_{p'} > 0$.

If the risk ratio or odds ratio improvement metrics are desired, Equations~\ref{eq:I_random} and \ref{eq:I_best} are modified accordingly but otherwise estimation is then carried out the same as in the previous section. \\

\subsubsection{For a Survival Response}\label{subsec:I_scores_survival}

Survival responses differ in two substantive ways from continuous responses: (1) they are always non-negative (2) some values are \qu{censored} which means it assumes the value of the last known measurement but it is certain that the true value is greater. The responses $\y$ are coupled with this censoring information $\c$, a binary vector of length $n$ where the convention is to let $c_i = 0$ to indicate that $y_i$ is censored and thus set equal to its last known value.

To obtain $\dhat$, we require a survival model. For example purposes here we will assume the exponential regression model (the exponentiation enforces the positivity of the response values) with the usual first order treatment interactions,

\bneqn\label{eq:exp_model}
f(\x_{i1}, A_i) = \text{exp}\parens{\beta_0 + \beta_{1}x_{1} + \ldots + \beta_{p}x_{p} + A_i \parens{\gamma_{0} + \gamma_{1'} x_{1'} + \ldots + \gamma_{p'}x_{p'}}}.
\eneqn

\noindent Under the exponential model, the convention is that the noise term $\errorrv$ is multiplicative instead of additive (i.e. $Y_i = f(\x_{i1}, A_i) \errorrv_i$). Note that at this step, a fully parametric model is needed; the non-parametric Kaplan-Meier or the semi-parametric Cox proportion hazard model are insufficient as we need a means of explicitly estimating $\cexpe{Y}{X, A}$ for all values of $X$ and both values of $A$.

Moreso than for continuous and incidence endpoints, parameter estimation is dependent on the choice of error distribution. Following \citet{Hosmer1999}, a flexible model is to let $\natlog{\errorrv_1}, \ldots, \natlog{\errorrv_n} \iid$ $\text{Gumbel}(0, \sigsq)$, implying the popular Weibull model for survival and the default in our software. As the case previously, the user is free to choose whatever model they wish. The $\beta_j$'s, $\gamma_j$'s and the nuisance scale parameter $\sigsq$ are fit using maximum likelihood taking care to ensure the correct contributions of censored and uncensored values. Similar to the case of logistic regression, the likelihood function does not have a closed form solution and must be approximated numerically.

Some algebra demonstrates that the estimated decision rule is the same as those above, i.e. $\hat{d}[\fhat(\x)] = 1$ when $\hat{\gamma}_{0} + \hat{\gamma}_{1'} x_{1'} + \ldots +\hat{\gamma}_{p'}x_{p'} > 0$. In other words, the subject is given the treatment that yields the longest expected survival. 

Subjects are then sorted in cells like Table~\ref{tab:two_by_two} but care is taken to keep the corresponding $c_i$ values together with their paired $y_i$ values, following \citet{Yakovlev1994}. At this point, we need to specify analogous computations to Equations~\ref{eq:I_random} and \ref{eq:I_best} that are sensitive to the fact that many $y_i$ values are censored. (The sample averages $\ybar$ obviously cannot be employed here because it ignores this censoring).

Of course we can reemploy a new Weibull model and define improvement as we did earlier as the difference in expectations (Equation~\ref{eq:mean_improvement_score}). However, there are no more covariates needed at this step as all subjects have been sorted based on $\dhat(\x)$. Thus, there is no reason to require a parametric model that may be arbitrarily wrong.

For our default implementation, we have chosen to employ the difference of the Kaplan-Meier median survival statistics here because we intuitively feel that a non-parametric estimate makes the most sense. Once again, the user is free to employ whatever they feel is most appropriate in their context. Given this default, please note that the improvement measure of Equation~\ref{eq:mean_improvement_score} is no longer defined as the difference in survival expectations, but now the difference in survival \emph{medians}. This makes our framework slightly different in the case of survival endpoints.\\

\subsection{Inference for the population improvement parameter}\label{subsec:cis_tests_I_measure}

The $\Ihato$ estimates are draws from an elaborate estimator whose sampling distribution is not available in closed form. We can employ the nonparametric bootstrap to obtain an asymptotic estimate of its sampling variability, which can be used to construct confidence intervals and testing procedures \citep{Efron1994}. 

In the context of our proposed methodology, the bootstrap procedure works as follows for the target of inference $\muIo$. We take a sample with replacement from the RCT data of size $n$ denoted with tildes: $[\tilde{\X}, \tilde{\y}]$. Using the 10-fold CV procedure described at the end of Section~\ref{subsec:I_scores}, we create an estimate $\tilde{{I}}_{\text{0}}$. We repeat the resampling of the RCT data and the recomputation of $\tilde{{I}}_{\text{0}}$ $B$ times where $B$ is selected for resolution of the confidence interval and significance level of the test. In practice we found $B=3000$ to be sufficient, so we leave this as the default in our software implementation. Because the $n$'s of usual RCT's are small, and the bootstrap is embarrassingly parallel, this is not an undue computational burden.

In this application, the bootstrap approximates the sampling of many RCT datasets. Each $\tilde{I}$ that is computed corresponds to one out-of-sample improvement estimate for a particular RCT dataset drawn from the population of RCT datasets. We stress again that the frequentist confidence intervals and tests that we develop for the improvement measure do \textit{not} constitute inference for a new \textit{individual's} improvement, it is inference for the average improvement for future subjects versus random allocation, $\muIo$.

To create an $1 - \alpha$ level confidence interval, first sort the $\{\tilde{{I}}_{\text{0}, 1}, \ldots, \tilde{{I}}_{\text{0}, B}\}$ by value, and then report the values corresponding to the empirical $\alpha / 2$ and $1 - \alpha / 2$ percentiles. This is called the \qu{percentile method}. \qu{Although this direct equation of quantiles of the bootstrap sampling distribution with confidence limits may seem initially appealing, it’s rationale is somewhat obscure} \citep[page 272]{Rice1994}. There are other ways to generate asymptotically valid confidence intervals using bootstrap samples but some debate about which has the best finite sample properties. We have also implemented the the \qu{basic method} \citep[page 194]{Davison1997} and the bias-corrected \qu{$BC_a$ method} of \citet{Efron1987} that \citet{DiCiccio1996} claim performs an order of magnitude better in accuracy than the percentile method. Implementing other confidence interval methods for the bootstrap may be useful future work.

If a higher response is better for the subject, we set $H_0: \muIo \leq 0$ and $H_a: \muIo > 0$. Thus, we wish to reject the null hypothesis that our allocation procedure is at most as useful as a naive business-as-usual procedure. To obtain an asymptotic $p$ value based on the percentile method, we count the number of bootstrap sample $\tilde{I}$ estimates below 0 and divide by $B$. This bootstrap procedure is graphically illustrated in the bottom half of Fig~\ref{fig:method_illustration} and the bootstrap confidence interval and $p$ value computation is illustrated in the center. Note that for incidence outcomes where the improvement is defined as the risk ratio or odds ratio, we use $H_0: \muIo \leq 1$ and $H_a: \muIo > 1$ and count the number of $\tilde{I}$ estimates below 1.

We would like to stress once again that we are not testing for \textit{qualitative interactions} --- the ability of a covariate to \qu{flip} the optimal treatment for subjects. Tests for such interactions would be hypothesis tests on the $\gamma$ parameters of Equations~\ref{eq:provided_model}, \ref{eq:logit_model} and \ref{eq:exp_model}, model structures that are not even required for our procedure. Qualitative interactions are controversial and entire tests have been developed to investigate their significance. In the beginning of Section~\ref{sec:background} we commented that most RCT's are not even powered to investigate these interactions. \qu{Even if an optimal personalized medicine rule [based on such interactions] can provide substantial gains it may be difficult to estimate this rule with few subjects} \citep{Rubin2012}. The bootstrap test (and our approach at large) looks at the holistic picture of the personalization scheme without focus on individual covariate-treatment interaction effects to determine if the personalization scheme in totality is useful, conceptually akin to the omnibus F-test in OLS.\\

\subsubsection{Concerns with using the bootstrap for this inference}

There is some concern in the personalized medicine literature about the use of the bootstrap to provide inference. First, the estimator for $V$ is a non-smooth functional of the data which may result in an inconsistent bootstrap estimator \citep{Shao1994}. The non-smoothness is due to the indicator function in Equation~\ref{value_estimate_non_smooth} being non-differentiable, similar to the example found in \citep[Section 4.3.1]{Horowitz2001}. However, \qu{the value of a fixed [response model] (i.e., one that is not data-driven) does not suffer from these issues and has been addressed by numerous authors} \citep{Chakraborty2014a}. Since our $\hat{V}$ is constructed out-of-sample, it is merely a difference of sample averages of the hold-out response values that are considered pre-sorted according to a fixed rule.\footnote{Note also that we do not have the additional non-smoothness created by Q-learning during the maximization step \citep[Section 2.4]{Chakraborty2009}.} This does not come without a substantial cost. Estimation of the improvement score out-of-sample means the effective sample size of our estimate is small and our power commensurately suffers. One can also implement the double bootstrap (see e.g. the comparisons in \citealp{Chakraborty2009}) herein and that is forthcoming in our software (see Section~\ref{sec:software}).

There is an additional concern. Some bootstrap samples produce null sets for the \qu{lucky subjects} (i.e. $P \cup S = \varnothing$ of Table~\ref{tab:two_by_two} or equivalently, all values of the indicator in Equation~\ref{value_estimate_non_smooth} are zero). These are safe to ignore as we are only interested in the distribution of estimates conditional on feasibility of estimation. Empirically, we have noticed as long as $n > 20$, there are less than 1\% of bootstrap samples that exhibit this behavior. Either way, we print out this percentage when using the \pkg{PTE} package and large percentages warn the user that the inference is suspect. \\

\subsection{Future Subjects}\label{subsec:future_subjects}

The implementation of this procedure for future patients is straightforward. Using the RCT data, estimate $f$ to arrive at $\fhat$. When a new individual, whose covariates denoted $\x_*$, enters a clinic, our estimated decision rule is calculated by predicting the response under both treatments, then allocating the treatment which corresponds to the better outcome, i.e. $\dhat(\x_*)$. This final step is graphically illustrated in the top left of Fig~\ref{fig:method_illustration}.

It is important to note that $\dhat(\x_*)$ is built with RCT data where treatment was allocated randomly and without regard to the subject covariates. In the example of the first order linear model with treatment interactions, the $\gamma$ parameters have a causal interpretation --- conditional causation based on the values of the moderating covariates. Thus $\dhat(\x_*)$ reflects a treatment allocation that \emph{causes} the response to be higher (or lower). We reiterate that this would not be possible with observational data which would suffer from elaborate confounding relationships between the treatment and subject covariates (see discussion in Sections~\ref{sec:background} and \ref{subsec:future_directions}).

\section{Data Examples}\label{sec:data}

We present two simulations in Sections~\ref{subsec:simulation_simple} and \ref{subsec:simulation_complicated} that serve only as illustrations that our methodology both works as purported but degrades in the case of pertinent information that goes missing. We then demonstrate a real clinical setting in Section~\ref{subsec:clinical_trial_example}.\\

\subsection{Simulation with correct regression model}\label{subsec:simulation_simple}

Consider a simulated RCT dataset with one covariate $x$ where the true response function is known:

\bneqn\label{eq:simple_model}
Y = \beta_0 + \beta_1 X + A(\gamma_0 + \gamma_1 X) + \errorrv
\eneqn

\noindent where $\errorrv$ is mean-centered. We employ $f(x, A)$ as the true response function, $\cexpe{Y}{X, A}$. Thus, $d = d^*$, the \qu{optimal} rule in the sense that a practitioner can make optimal allocation decisions (modulo noise) using $d(x) = \indic{\gamma_0 + \gamma_1 x > 0}$. Consider $d_0$ to be the \textbf{random} allocation procedure (see Section~\ref{subsec:naive_allocation_procedures}). Note that within the improvement score definition (Equation~\ref{eq:mean_improvement_score}), the notation $\expesubsup{\X}{d}{Y}$ is an expectation over the noise $\errorrv$ and the joint distribution of $X,~A$. After taking the expectation over noise, the improvement under the model of Equation~\ref{eq:simple_model} becomes

\beqn
\muIo &=& \expesub{X}{\beta_0 + \beta_1 X + \indic{\gamma_0 + \gamma_1 X > 0} (\gamma_0 + \gamma_1 X)} - \expesub{X}{\beta_0 + \beta_1 X + 0.5 (\gamma_0 + \gamma_1 X)} \\
&=& \expesub{X}{\parens{\indic{\gamma_0 + \gamma_1 x > 0} - 0.5}\parens{\gamma_0 + \gamma_1 X}} \\
&=& \gamma_0 \parens{\prob{\gamma_0 + \gamma_1 X > 0} - 0.5} + \gamma_1 \parens{\expesub{X}{X\indic{\gamma_0 + \gamma_1 x > 0}} -0.5 \expesub{X}{X}}. \\
\eeqn

\noindent We further assume $X \sim \normnot{\mu_X}{\sigsq_X}$ and we arrive at

\beqn 
\muIo = (\gamma_0 + \gamma_1\mu_X) \parens{.5 - \Phi\parens{-\frac{\gamma_0}{ \gamma_1}}} + \gamma_1 \dfrac{\sigma_X}{\sqrt{2\pi}}~\exp{-\oneover{2\sigsq_X}{\parens{-\frac{\gamma_0}{\gamma_1}-\mu}^2}}.
\eeqn

\begin{figure}[t]
\centering
\includegraphics[width=5in]{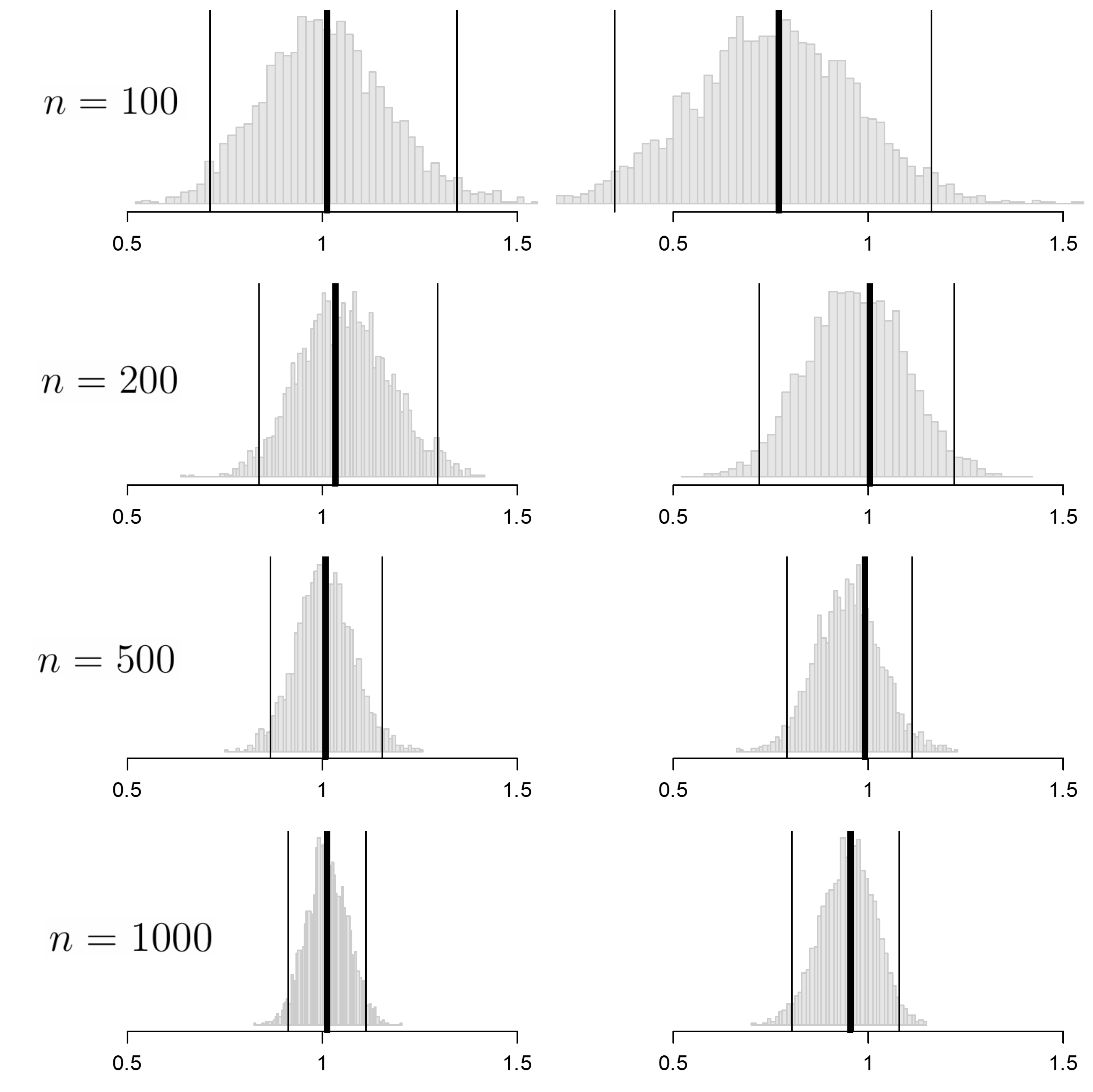}
\caption{Histograms of the bootstrap samples of the out-of-sample improvement measures for $d_0$ \textbf{random} (left column) and $d_0$ \textbf{best} (right column) for the response model of Equation~\ref{eq:simple_model} for different values of $n$. $\Ihato$ is illustrated with a thick black line. The $CI_{\muIo, 95\%}$ computed by the percentile method is illustrated by thin black lines.}
\label{fig:simple_sim_results}
\end{figure}

We simulate under a simple scenario to clearly highlight features of our methodology. If $\mu_X = 0,~\sigsq_X = 1$ and $\gamma_0 = 0$, neither treatment $T_1$ or $T_2$ is on average better. However, if $x > 0$, then treatment $T_2$ is better in expectation by $\gamma_1 \times x$ and analogously if $x < 0$, $T_1$ is better by $-\gamma_1 \times x$. We then set $\gamma_1 = \sqrt{2\pi}$ to arrive at the round number $\muIo = 1$. We set $\beta_0 = 1$ and $\beta_1 = -1$ and let $\errorrv_i \iid \stdnormnot$. We let the treatment allocation vector $A$ be a random block permutation of size $n$, balanced between $T_1$ and $T_2$. Since there is no PATE, the \textbf{random} and \textbf{best} $d_0$ procedures (see Section~\ref{subsec:naive_allocation_procedures}) are the same in value. We then vary $n \in \braces{100, 200, 500, 1000}$ to assess convergence for both $d_0$ procedures and display the results in Fig~\ref{fig:simple_sim_results}.

Convergence to $\muIo = 1$ is observed clearly for both procedures but convergence for $d_0$ \textbf{best} is slower than $d_0$ \textbf{rand}. This is due to the $\Vhat$ being computed with fewer samples: $\ybar_{\text{test}}$, which uses all of the available data, versus $\ybar_{P \cup Q}$ or $\ybar_{R \cup S}$, which uses only half the available data on average (see Equations~\ref{eq:I_rand} and \ref{eq:I_best}) Also note that upon visual inspection, our bootstrap distributions seem to be normal. Non-normality in this distribution when using the software package warns the user that the inference is suspect.

In this section we assumed knowledge of $f$ and thereby had access to an optimal rule. In the next section we explore convergence when we do not know $f$ but pick an approximate model yielding a non-optimal rule.

\subsection{Simulation with an approximate regression model}\label{subsec:simulation_complicated}

Consider RCT data with a continuous endpoint where the true response model is

\bneqn\label{eq:complicated_sim_model}
Y = \beta_0 + \beta_1 X + \beta_2 U + A (\gamma_0 + \gamma_1 X^3 + \gamma_2 U) + \errorrv 
\eneqn

\noindent where $X$ denotes a covariate recorded in the RCT and $U$ denotes a covariate that is not included in the RCT dataset. The optimal allocation rule $d^*$ is 1 when $\gamma_0 + \gamma_1 X^3 + \gamma_2 U > 0$ and 0 otherwise. The practitioner, however, does not have access to the information contained in $U$, the unobserved covariate, and has no way to ascertain the exact relationship between $X$ and the treatment. Consider the default model that is an approximation of the true population response surface,

\bneqn\label{eq:complicated_sim_model_approx}
f(X, A) = \beta_0 + \beta_1 X + A (\gamma_0 + \gamma_1 X),
\eneqn

\noindent which is different from the true response model due to (a) the misspecification of $X$ (linear instead of cubic) and (b) the absence of covariate $U$ (see Equation~\ref{eq:response_model_decomposition}). This is the more realistic scenario; even with infinite data, $d^*$ cannot be located because of both ignorance of the true model form and unmeasured subject characteristics.

To simulate, we set the $X$'s, $U$'s and $\errorrv$'s to be standard normal variables and then set $\beta_0 = 1, ~ \beta_1 = -1, ~ \beta_2 = 0.5, ~ \gamma_0 = 0, ~\gamma_1 = 1$ and $\gamma_2 = -3$. The $X_i$'s and the $U_i$'s are deliberately made independent of one another so that the observed covariates cannot compensate for the unobserved covariates, making the comparison between the improvement under $d^*$ and $d$ more stark. To find the improvement when the true model's $d^*$ is used to allocate, we simulate under Equation~\ref{eq:complicated_sim_model} and obtain $\muIo^* \approx 1.65$ and analogously, to find the improvement under approximation model's $d$, we simulate under Equation~\ref{eq:complicated_sim_model_approx} and obtain $\muIo \approx 0.79$. Further simulation shows that not observing $U$ is responsible for 85\% of this observed drop in performance and employing the linear $X$ in place of the non-linear $X^3$ is responsible for the remaining 15\%. Since $\gamma_0 = 0$ and the seen and unseen covariates are mean-centered, there is no PATE and thus these simulated improvements apply to both the cases where $d_0$ is \textbf{random} and $d_0$ is \textbf{best}.

\begin{figure}[t]
\centering
\includegraphics[width=5in]{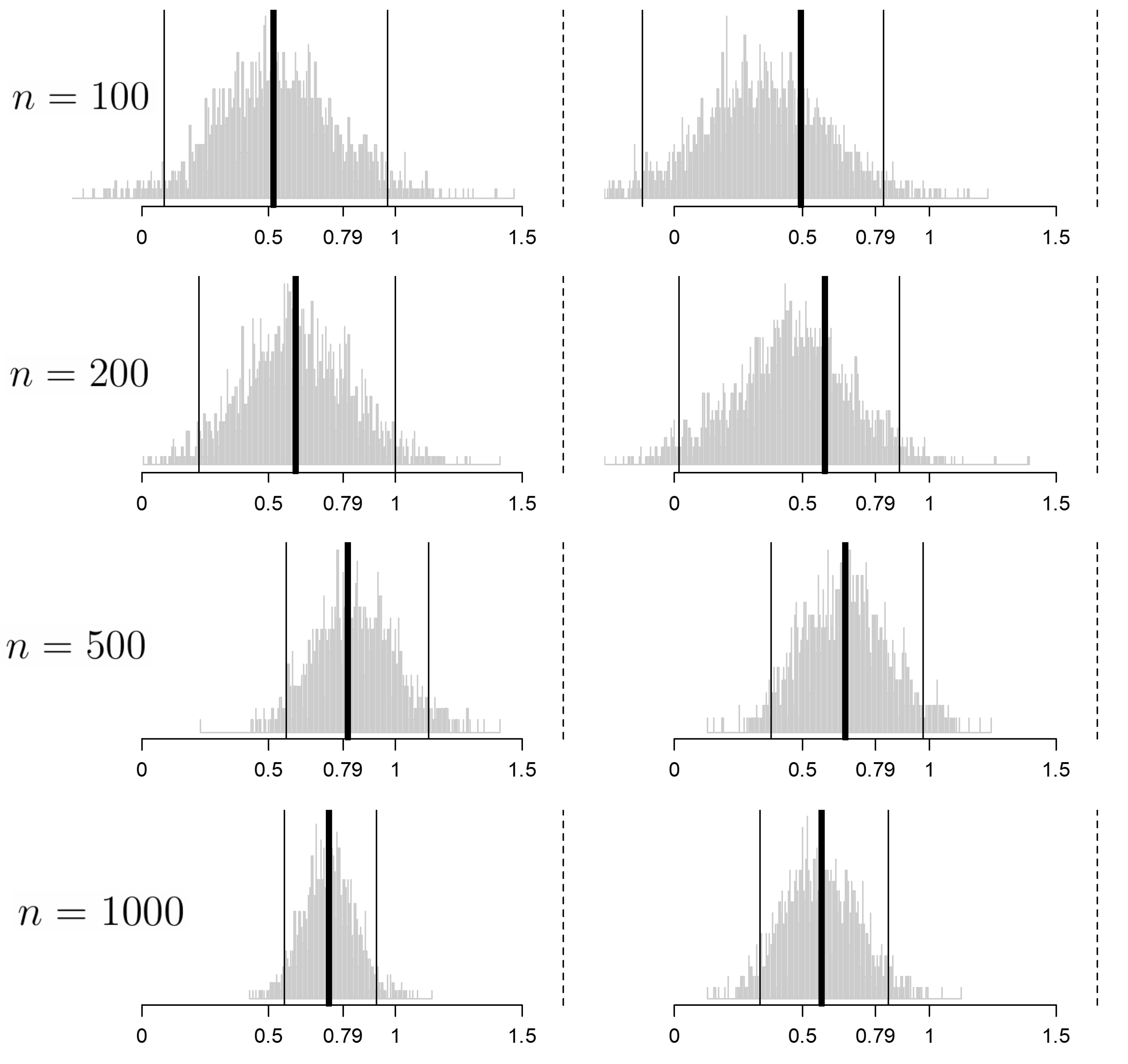}
\caption{Histograms of the bootstrap samples of the cross-validated improvement measures for $d_0$ \textbf{random} (left column) and $d_0$ \textbf{best} (right column) for the response model of Equation~\ref{eq:complicated_sim_model} for different values of $n$. $\Ihato$ is illustrated with a thick black line. The $CI_{\muIo, 95\%}$ computed via the percentile method is illustrated by thin black lines. The true population improvement $\muIo^*$ given the optimal rule $d^*$ is illustrated with a dotted black line}
\label{fig:complicated_sim_results}
\end{figure}

Fig~\ref{fig:complicated_sim_results} demonstrates results for $n = \braces{100, 200, 500, 1000}$ analogous to Fig~\ref{fig:simple_sim_results}. We observe that the bootstrap confidence intervals contain $\muIo$ but not $\muIo^*$. This is expected; we are not allocating using an estimate of $d^*$, only an estimate of $d$.

Convergence towards $\muIo = 0.79$ is observed clearly for both procedures and once again the convergence is slower for the \textbf{best} procedure for the same reasons outlined in Section~\ref{subsec:simulation_simple}. Note that the coverage illustrated here is far from $\muIo^*$, the improvement using the optimal allocation rule. \citet{Kallus2017} presents a coefficient of personalization metric similar to $R^2$ where a value of 100\% represents perfect personalization and 0\% represents standard of care. Here, we would fall far short of the 100\%.

The point of this section is to illustrate what happens in the real world: the response model is unknown and important measurements are missing and thus any personalized medicine model falls far short of optimal. However, the effort can still yield an improvement that can be clinically significant and useful in practice.

There are many cases where our procedure will not find signal yielding improvement in patient outcomes. For example (1) in cases where there are many variables that are important and a small sample size. Clinical trials are not usually powered to find even single interaction effects, let alone many. The small sample size diminishes power to find the effect, similar to any statistical test. (2) If the true heterogeneity in the functional response cannot be approximated by a linear function. For instance, parabolic or sine functions cannot be represented whatsoever by best fit lines.

In the next section, we use our procedure in RCT data from a real clinical trial where both these limitations apply. The strategy is to approximate the response function using a reasonable model $f$ built from domain knowledge and the variables at hand and hope to find demonstrate a positive, clinically meaningful $\muIo$ knowing full well it will be much smaller than $\muIo^*$.

\subsection{Clinical trial demonstration}\label{subsec:clinical_trial_example}

We consider the RCT data of \citet{DeRubeis2005} where there are two depression treatments with very different action mechanisms: cognitive behavioral therapy ($T_1$) and an antidepressant medication paroxetine ($T_2$). After omitting patients who dropped out there were $n = 154$ subjects with 28 baseline characteristics measured. Although this dataset did not have explicit patient-level identifying information, using these 28 characteristics could potentially identify some of the patients. Note that this study was funded and begun before clinical trials registration and thus it does not have a clinical trial registration number.

The primary outcome measure $y$ is the continuous Hamilton Rating Scale for Depression (HRSD), a composite score of depression symptoms where lower means less depressed, assessed by a clinician after 16 weeks of treatment. A simple $t$ test revealed that there was no statistically significant difference between the cognitive behavioral therapy and paroxetine, a well-supported finding in the depression literature. Despite the seeming lack of a PATE, practitioner intuition and a host of studies suggest that the covariates collected can be used to build a principled personalized model with a significant negative $\muIo$. The lack of a PATE also suggests that the \textbf{random} $d_0$ is an appropriate baseline comparison.

We now must specify a model, $f$. For the purpose of illustration, we employ a linear model with first-order interactions with the treatment (as in Equation~\ref{eq:provided_model}). Which of the 28 variables should be included in the model? Clinical experience and theory should suggest both mediator (main effect) and moderator (treatment interaction) variables \citep{Cohen2017}. We should not use variables selected using methods performed on this RCT data such as the variables found in \citet[Table 3]{DeRubeis2014}. Such a procedure would constitute data snooping and it will invalidate the inference provided by our method. The degree of invalidation is not currently known and is much needed to be researched.

Of the characteristics measured in this RCT data, previous researchers have found significant treatment moderation in age and chronicity \citep{Cuijpers2012}, early life trauma \citep{Nemeroff2003} (which we approximate using a life stressor metric), presence of personality disorder \citep{Bagby2008}, employment status and marital status \citep{Fournier2009} but almost remarkably baseline severity of the depression does not moderate \citep{Weitz2015}. We include these $p=6$ variables as moderators and as mediators.\footnote{Although this is standard modeling practice, it is not absolutely essential in our methodology, where our goal is neither inference for the variables nor prediction of the endpoint.}

%baseline HRSD score, IQ, age, married (or not), presence of chronic depression, a measure of life stressors, drug); the treatment moderating variables were marital status, employment status, degree of life stressors, personality disorder and whether the patient was taking other drugs.

%When fitting a linear model to capture this theory, .  The next Section will demonstrate how a model such as this one is entered into our software and fit with least squares to generate $\hat{d}$.

Using out bootstrap inference methodology with $B=3000$, we estimate the improvement of this model, confidence intervals and their statistical significances. We anticipate that a new subject allocated using this personalization model will be less depressed on average by 1.16 HRSD units with a 95\% percentile confidence interval of $\bracks{-2.68, -0.30}$ compared to that same subject being allocated randomly to cognitive behavioral therapy or paroxetine. We can reject the null hypothesis that personalization is no better than random allocation for a new subject ($p$ value = 0.008). 

In short, the results are statistically significant, but the estimated improvement may not be clinical significant. According to the criterion set out by the National Institute for Health and Care Excellence, three points on the HRSD is considered clinically important. Nevertheless, this personalization scheme can be implemented in practice with new patients for a modest improvement in patient outcome at little cost.

\section[The PTE Package]{The \pkg{PTE} Package}\label{sec:software}

%We demonstrate basic syntax of the \pkg{PTE} package software here employing example code that can be emulated. For more specific details on the software's options, see the help pages from within \proglang{R}. 

\subsection{Estimation and Inference for Continuous Outcomes}

The package comes with two example datasets. The first is the continuous data example. Below we load the library and data.

\begin{verbatim}
R> library(PTE); library(dplyr)
R> data(continuous_example)
R> X = continuous_example$X
R> y = continuous_example$y
R> continuous_example$X %>% sample_n(5)
# A tibble: 5 x 6
  treatment     x1     x2     x3     x4         x5
      <dbl> <fctr> <fctr> <fctr> <fctr>      <dbl>
1         1     NO    OFF    YES MEDIUM  1.3009448
2         0    YES    OFF    YES MEDIUM -0.5483983
3         0     NO    OFF    YES    LOW  0.3762733
4         1     NO    OFF    YES MEDIUM -1.1648459
5         1    YES     ON    YES   HIGH -0.8566221
> round(head(continuous_example$y), 3)
[1] -0.746 -1.359  0.020  0.632 -0.823 -2.508
\end{verbatim}

The endpoint $y$ is continuous and the RCT data has a binary treatment vector appropriately named (this is required) and five covariates, four of which are factors and one is continuous. We can run the estimation for the improvement score detailed in Section~\ref{subsec:I_scores_continuous} and the inference of Section~\ref{subsec:cis_tests_I_measure} by running the following code:

\begin{verbatim}
R> options(mc.cores = 4)
R> pte_results = PTE_bootstrap_inference(X, y, 
    B = 1000, run_bca_bootstrap = TRUE)
R> plot(pte_results)
\end{verbatim}

\noindent Here, 1000 bootstrap samples were run on four cores in parallel to minimize runtime. The model defaults to a linear model where all variables included are interacted with the treatment and fit with least squares. Below are the results.

\begin{verbatim}
R> pte_results
    I_random observed_est = 0.077,  pctile p-val = 0.021, 
      95% CI's: basic = [-0.237, 0.143], pctile = [0.011, 0.391], 
        BCa = [-0.085, 0.145],
    I_best observed_est = 0.065,  pctile p-val = 0.089, 
      95% CI's: basic = [-0.198, 0.2], pctile = [-0.071, 0.328], 
        BCa = [-0.146, 0.198]
\end{verbatim}

Note how the three bootstrap methods are different from another. The percentile method barely includes the actual observed statistic for the random comparison (see discussion in Section~\ref{subsec:cis_tests_I_measure}). The software also plots the $\tilde{I}$'s in a histogram (unshown).

To demonstrate the flexibility of the software, consider the case where the user wishes to use $x_1, x_2, x_3, x_4$ as mediators and $x_5$ as the sole treatment moderator. And further, the user wishes to estimate the model parameters using the ridge penalty instead of OLS. Note that this is an elaborate model that would be difficult to justify in practice and it is only shown here as an illustration of the customizability of the software. Below is the code used to test this approach to personalization.

\begin{verbatim}
R> library(glmnet)
R> pte_results = PTE_bootstrap_inference(X, y, B = 1000, 
  personalized_model_build_function = function(Xytrain){    
    Xytrain_mm = model.matrix(~ . - y + x5 * treatment, Xytrain)
    cv.glmnet(Xytrain_mm, Xytrain[, ncol(Xytrain)], alpha = 0)
  },
  predict_function = function(mod, Xyleftout){
    Xyleftout$censored = NULL
    Xyleftout_mm = model.matrix(~ . + x5 * treatment, Xyleftout)
    predict(mod, Xyleftout_mm)
  })
\end{verbatim}

\noindent Here, the user passes in a custom function that builds the ridge model to the argument \code{personalized\_model\_build\_function}. The specification for ridge employed here uses the package \pkg{glmnet} \citep{Friedman2010} that picks the optimal ridge penalty hyperparameter automatically. Unfortunately, there is added complexity: the \pkg{glmnet} package does not accept formula objects and thus model matrices are generated both upon model construction and during prediction. This is the reason why a custom function is also passed in via the argument \code{predict\_function} which wraps the default \pkg{glmnet} \code{predict} function by passing in the model matrix.\\

\subsection{Estimation and Inference for Binary Outcomes}

In order to demonstrate our software for the incidence outcome, we use the previous data but threshold its response arbitrarily at its 75\%ile to create a mock binary response (for illustration purposes only). 

\begin{verbatim}
R> y = ifelse(y > quantile(y, 0.75), 1, 0)
\end{verbatim}

\noindent We then fit a linear logistic model using all variables as fixed effects and interaction effects with the treatment. As discussed in Section~\ref{subsec:I_scores_binary}, there are three improvement metrics for incidence outcomes. The default is the odds ratio. The following code fits the model and performs the inference.

\begin{verbatim}
R> pte_results = PTE_bootstrap_inference(X, y, B = 1000, 
    regression_type = "incidence", run_bca_bootstrap = TRUE)
Warning message:
glm.fit: fitted probabilities numerically 0 or 1 occurred
\end{verbatim}

\noindent Note that the response type \code{incidence} has to be explicitly made known otherwise the default would be regression. Below are the results.

\begin{verbatim}
R> pte_results
    I_random observed_est = 1.155,  pctile p-val = 0.104, 
      95% CI's: basic = [0.317, 1.475], pctile = [0.836, 1.994], 
        BCa = [0.497, 1.503],
    I_best observed_est = 1.04,  pctile p-val = 0.323, 
      95% CI's: basic = [0.329, 1.431], pctile = [0.649, 1.751], 
        BCa = [0.52, 1.528]
\end{verbatim}

\noindent The $p$ value is automatically calculated for $H_0: \mu_{I_0} < 1$ (i.e. the odds of improvement is better in $d_0$ than $d$). Other tests can be specified by changing the \code{H\_0\_mu\_equals} argument. Here, the test failed to reject $H_0$. Information is lost when a continuous metric is coerced to be binary. If the user wished to define improvement via the risk ratio (or straight probability difference), an argument would be added to the above, \code{incidence\_metric = "risk\_ratio"} (or \code{"probability\_difference"}).\\

\subsection{Estimation and Inference for Survival Outcomes}

Our package also comes with a mock RCT dataset with a survival outcome. Below, we load the data.

\begin{verbatim}
R> data(survival_example)
R> X = survival_example$X
R> y = survival_example$y
R> censored = survival_example$censored
\end{verbatim}

\noindent There are four covariates, one factor and three continuous. We can run the estimation for the improvement score detailed in Section~\ref{subsec:I_scores_survival} and inference for the true improvement by running the following code.

\begin{verbatim}
R> pte_results = PTE_bootstrap_inference(X, y, censored = censored, 
  B = 1000, run_bca_bootstrap = TRUE, regression_type = "survival")
\end{verbatim}

The syntax is the same as the above two examples except here we pass in the binary $\c$ vector separately and declare that the endpoint type is survival. Again by default all covariates are included as main effects and interactions with the treatment in a linear Weibull model. 

In the default implementation for the survival outcome, improvement is defined as median survival difference of personalization versus standard of care. The median difference can be changed via the user passing in a new function with the \code{difference\_function} argument. The median difference results are below.

\begin{verbatim}
R> pte_results
    I_random observed_est = 0.148,  p val = 0.017, 
      95% CI's: basic = [0.011, 0.296], pctile = [0, 0.285], 
        BCa = [-0.001, 0.274],
    I_best observed_est = -0.041,  p val = 0.669, 
      95% CI's: basic = [-0.134, 0.082], pctile = [-0.164, 0.052], 
        BCa = [-0.195, 0.013]
\end{verbatim}

It seems that the personalized medicine model increases median survival by 0.148 versus $d_0$ being the random allocation of the two treatments. If survival was measured in years (the typical unit), this would be about 2 months. However, it cannot beat the $d_0$ being the \textbf{best} of the two treatments. Remember, this is a much more difficult improvement metric to estimate as we are really comparing two cells in Table~\ref{tab:two_by_two} to another two cells, one of which is shared. Thus the sample size is low and power suffers. This difficulty is further compounded in the survival case because censored observations add little information.

\section{Discussion}\label{sec:discussion}

We have provided a methodology to test the effectiveness of personalized medicine models. Our approach combines RCT data with a statistical model $f$ of the response for estimating \textit{improved} outcomes under different treatment allocation protocols. Using the non-parametric bootstrap and cross-validation, we are able to provide confidence bounds for the improvement and hypothesis tests for whether the personalization performs better compared to a business-as-usual procedure. We demonstrate the method's performance on simulated data and on data from a clinical trial on depression. We also present our statistical methods in an open source software package in \proglang{R} named \pkg{PTE} which is available on \texttt{CRAN}.\\

\subsection{Limitations and Future Directions}\label{subsec:future_directions}

Our method and corresponding software have been developed for a particular kind of RCT design. The RCT must have two arms and one endpoint (continuous, incidence or survival). An extension to more than two treatment arms is trivial as Equation~\ref{eq:d} is already defined generally. Implementing extensions to longitudinal or panel data are simple within the scope described herein. And extending the methodology to count endpoints would also be simple.

Although we agree that \qu{a `once and for all' treatment strategy [may be] suboptimal due to its inflexibility}  \citep{Zhao2015}, this one-stage treatment situation is still common in the literature and the real world and this is the setting we chose to make our impact. We consider an extended implementation for dynamic treatment regimes on multi-stage experiments fruitful future work. Consider being provided with RCT data from sequential multiple assignment randomized trials (\qu{SMARTs}, \citealp{Murphy2005a}) and an a priori response model $f$. The estimate of $\Vhat[\dhat]$ (Equation~\ref{eq:estimated_improvement_score}) can be updated for a SMART with $k$ stages \citep{Chakraborty2014a} where our Table~\ref{tab:two_by_two} is a summary for only a single stage. In a SMART with $k$ stages, the matrix becomes a hypercube of dimension $k$. Thus, the average of diagonal entries in the multi-dimensional matrix is the generalization of the estimate of $\Vhat[\dhat]$ found in Equation~\ref{eq:I_rand}. Many of the models for dynamic treatment regimes found in \citet{Chakraborty2013} can then be incorporated into our methodology as $d$, and we may be able to provide many of these models with valid statistical inference. Other statistics computed from this multi-dimensional matrix may be generalized as well.

Our choices of $d_0$ explored herein were limited to the \textbf{random} or the \textbf{best} procedures (see Section~\ref{subsec:naive_allocation_procedures}). There may be other business-as-usual allocation procedures to use here that make for more realistic baseline comparisons. For instance, one can modify \textbf{best} to only use the better treatment if a two-sample t-test rejects at prespecified Type I error level and otherwise default to \textbf{random}. One can further set $d_0$ to be a regression model or a physician's decision tree model and then use our framework to pit two models against each other.

It might also be useful to consider how to extend our methodology to observational data. The literature reviewed in Section~\ref{sec:background} generally does not require RCT data but ``only'' a model that accurately captures selection into treatments e.g. if \qu{the [electronic medical record] contained all the patient information used by a doctor to prescribe treatment up to the vagaries and idiosyncrasies of individual doctors or hospitals} \citep[Section 1]{Kallus2017}. This may be a very demanding requirement in practice. In this paper, we do not even require valid estimates of the true population response surface. In an observational study one would need that selection model to be correct and/or a correct model of the way in which subjects and treatments were paired \citep{Freedman2008b}. Although assuming one has a model that captures selection, it would be fairly straightforward to update the estimators of Section~\ref{subsec:I_scores} to inverse weight by the probability of treatment condition (the \qu{IPWE}) making inference possible for observational data \citep{Zhang2012, Chakraborty2014a, Kallus2017}.

Another extension would be to drop the requirement of specifying the model $f$. This is a tremendous constraint in practice: what if the practitioner cannot construct a suitable $f$ using domain knowledge and past research? It is tempting to use a machine learning model that will both specify the structure of $f$ and provide parameter estimates within e.g. Kallus's personalization forests \citep{Kallus2017}. We believe the bootstrap of Section~\ref{subsec:cis_tests_I_measure} will withstand such a machination but are awaiting a rigorous proof. Is there a solution in the interim? As suggested as early as \citet{Cox1975}, we can always pre-split the data in two where the first piece can be used to specify $f$ and the second piece can be injected into our procedure. The cost is less data for estimation and thus, less power available to prove that the personalization is effective. 

If we do not split, all the data is to be used and there are three scenarios that pose different technical problems. Under one scenario, a researcher is able to specify a suite of possible models before looking at the data. The full suite can be viewed as comprising a single procedure for which nonparametric bootstrap procedures may in principle provide simultaneous confidence intervals \citep{Buja2014}. Under the other two scenarios, models are developed inductively from the data. This problem is more acute for instance in \citet{Davies2015} where high-dimensional genomic data is incorporated for personalization (e.g. where there are many more SNPs than patients in the RCT). If it is possible to specify exactly how the model search us undertaken (e.g., using the lasso), some forms of statistical inference may be feasible. This is currently an active research area; for instance, \citet{Lockhart2013} and \citet{Lee2016} develop a significance test for the lasso and there is even some evidence to suggest that the double-peeking is not as problematic as the community has assumed \citep{Zhao2017}. 

Our method's generalizability to future patients is also in question as our validation was done within the patients of a RCT. The population of future patients is likely not the same as the population of patients in the RCT. Future patients will likely have wider distributions of the $p$ covariates as typical RCT's feature strict inclusion criteria sometimes targeting high risk patients for higher outcome event rates. A good discussion of these issues is found in \citet[Chapter 6]{Rosenberger2016}. The practitioner will have to draw on experience and employ their best judgment to decide if the estimates our methodology provides will generalize.

And of course, the method herein only evaluates if a personalization scheme works on average over an entire population. \qu{Personalized medicine} eponymously refers to personalization for an individual. That is not the goal herein, but we do acknowledge that estimates and inference at an individual level coupled to valid inference for the improvement score is much-needed. This is not without difficulty as clinical trials are typically not powered to examine subgroup effects. A particularly alarming observation is made by \citet[page 7]{Cuijpers2012}, \qu{if we want to have sufficient statistical power to find clinically relevant differential effect sizes of 0.5, we would need .... about 23,000 patients}.

\section*{Conflict of Interest Statement}

The authors declare that the research was conducted in the absence of any commercial or financial relationships that could be construed as a potential conflict of interest.

\section*{Author Contributions}

All authors were responsible for development of methodology and drafting the manuscript. Kapelner and Bleich did the data analysis of Section~\ref{sec:data}.

\section*{Funding}

Adam Kapelner acknowledges a PSC-CUNY grant.

\section*{Acknowledgments}

We would like to thank our three reviewers and the editor whose input greatly improved this work. We would like to thank Abba Krieger, Larry Brown, Bibhas Chakraborty, Andreas Buja, Ed George, Dan McCarthy, Linda Zhao and Wei-Yin Loh for helpful discussions and comments. Earlier versions of this manuscript were released as a preprint on arXiv \citep{Kapelner2014}.

\section*{Supplemental Data}

The results and simulations in this paper (for which the code was not expressly found herein) can be duplicated by running the \proglang{R} scripts found at \url{github.com/kapelner/PTE/tree/master/paper_duplication}. 

\section*{Data Availability Statement}

The datasets presented in this article in Section~\ref{subsec:clinical_trial_example} is not readily available because the clinical data cannot be anonymized sufficiently according to the IRB guidelines of the institutions where the study was performed. For more information, contact the authors of the study.
%\pagebreak

\bibliographystyle{apa}
\bibliography{better_personalized_medicine}

\end{document}

%% file: preamble.tex
\usepackage{graphicx}
\usepackage{subcaption}
\usepackage{amsmath}
\usepackage{dsfont}
\usepackage{placeins}
\usepackage{amssymb}
\usepackage{abstract}
\usepackage[auth-sc,affil-sl]{authblk}
\usepackage{hyperref}
\usepackage{apacite}
\usepackage[margin=1in]{geometry}
\usepackage{enumerate}
\usepackage{fancyhdr}
\usepackage[xcdraw,table]{xcolor}
\usepackage{natbib}
\usepackage{algorithm}
\usepackage{algorithmicx}
\usepackage{algcompatible}
\usepackage{algpseudocode}
\usepackage[table]{xcolor}
\usepackage{multirow}
\usepackage{amsthm}

\newcommand{\qu}[1]{``#1''}

\newcounter{probnum}
\allowdisplaybreaks

\DeclareMathOperator*{\argmax}{arg\,max~}

\newcommand{\bv}[1]{\boldsymbol{#1}}

\newcommand{\proglang}[1]{\texttt{#1}}
\newcommand{\pkg}[1]{\texttt{#1}}
\newcommand{\code}[1]{\texttt{#1}}

\newcommand{\sigsq}{\sigma^2}

\newcommand{\thetavec}{\bv{\theta}}

\newcommand{\ybar}{\bar{y}}

\newcommand{\iid}{~{\buildrel iid \over \sim}~}

\newcommand{\A}{\bv{A}}

\newcommand{\U}{\bv{U}}
\newcommand{\X}{\bv{X}}

\newcommand{\x}{\bv{x}}

\newcommand{\y}{\bv{y}}

\renewcommand{\c}{\bv{c}}

\newcommand{\reals}{\mathbb{R}}

\newcommand{\beqn}{\vspace{-0.25cm}\begin{eqnarray*}}
\newcommand{\eeqn}{\end{eqnarray*}}
\newcommand{\bneqn}{\vspace{-0.25cm}\begin{eqnarray}}
\newcommand{\eneqn}{\end{eqnarray}}

\newcommand{\parens}[1]{\left(#1\right)}

\newcommand{\prob}[1]{\mathbb{P}\parens{#1}}
\newcommand{\cprob}[2]{\prob{#1~|~#2}}

\newcommand{\bracks}[1]{\left[#1\right]}
\newcommand{\braces}[1]{\left\{#1\right\}}

\newcommand{\expe}[1]{\mathbb{E}\bracks{#1}}
\newcommand{\cexpe}[2]{\expe{#1 ~ | ~ #2}}

\newcommand{\expesub}[2]{\mathbb{E}_{#1}\bracks{#2}}
\newcommand{\expesubsup}[3]{\mathbb{E}_{#1}^{#2}\bracks{#3}}
\newcommand{\indic}[1]{\mathds{1}_{#1}}

\renewcommand{\exp}[1]{\mathrm{exp}\parens{#1}}

\newcommand{\natlog}[1]{\ln\parens{#1}}
\newcommand{\oneover}[1]{\frac{1}{#1}}

\newcommand{\normnot}[2]{\mathcal{N}\parens{#1,\,#2}}

\newcommand{\stdnormnot}{\normnot{0}{1}}

\newcommand{\errorrv}{\mathcal{E}}

\newcommand{\Xtrain}{\X_{\text{train}}}
\newcommand{\ytrain}{\y_{\text{train}}}
\newcommand{\Xtest}{\X_{\text{test}}}
\newcommand{\ytest}{\y_{\text{test}}}
\newcommand{\ntrain}{n_{\text{train}}}
\newcommand{\ntest}{n_{\text{test}}}
\newcommand{\xtesti}{\x_{\text{test}, i}}

\newcommand{\muIo}{\mu_{I_0}}

\newcommand{\Ihato}{\hat{I}_0}
\newcommand{\fhat}{\hat{f}}
\newcommand{\fhattrain}{\fhat_{\text{train}}}
\newcommand{\dhat}{\hat{d}}
\newcommand{\Vhat}{\hat{V}}